\documentclass[aps,pre,twocolumn,groupedaddress,showpacs]{revtex4}
\usepackage{amsmath,amssymb}
\usepackage{color}
\usepackage[dvipdfm]{hyperref}
\usepackage[final,dvips]{graphicx}

\begin{document}
\pagecolor{white}
\color{black}

\title{State selection in the noisy stabilized Kuramoto Sivashinsky equation}

\author{D. Obeid$^{1}$}
\email{dinaobeid@gmail.com}
\author{J.~M. Kosterlitz$^{1,3}$}
\email{mikekost@het.brown.edu}
\author{B. Sandstede$^{2}$}
\email{Bjorn\_Sandstede@brown.edu}
\affiliation{$^{1}$Department of Physics, Brown University, Providence, RI 02912, USA}
\affiliation{$^{2}$Division of Applied Mathematics, Brown University, Providence, RI 02912, USA}
\affiliation{$^{3}$School of Computational Sciences, Korea Institute for Advanced Study, 207-43 Cheongryangni-2Dong, Dongdaemun-gu, Seoul 130-722, Korea}
\date{\today}

\begin{abstract}
In this work, we study the $1D$ stabilized Kuramoto Sivashinsky equation with additive uncorrelated stochastic noise. The Eckhaus stable band of the deterministic equation collapses to a narrow region near the center of the band. This is consistent with the behavior of the phase diffusion constants of these states. Some connections to the phenomenon of state selection in driven out of equilibrium systems are made. 
\end{abstract}


\maketitle

\section{Introduction}
In this paper, the old question of pattern selection in extended systems is studied in a simple but non trivial driven out of equilibrium model with additive Gaussian distributed stochastic noise. A number of physical systems fall into this category, examples of which are wavelength selection in directional solidification \cite{theoraticalwork1} and Rayleigh-B\'enard convection \cite{Cross&Hohenberg}. The absence of selection is claimed in models of convection \cite{crosstesauro} and in directional solidification \cite{dombre,amar}. Contrary claims have also been made \cite{karma86,kerszberg,kurtze,filho}. The apparently simple model we choose to study is the stabilized Kuramoto Sivashinsky equation in one spatial dimension with additive uncorrelated Gaussian distributed stochastic noise. This equation has the essential ingredient of non linearity, has a band of stable steady states in the absence of noise, displays most of the instabilities and stationary states of real systems \cite{misbahvallance} and is sufficiently simple to allow a detailed analysis in the presence of stochastic noise. All real systems are subject to some sort of stochastic noise which may have very important consequences. For example, a system evolving towards equilibrium will ultimately end up in a unique stationary equilibrium state which is the state of minimum free energy. There may be several states corresponding to local free energy minima but the system eventually reaches the state of absolute minimum free energy. Should the system evolve in a noiseless deterministic fashion, it will become stuck in the first local minimum encountered which implies that its final state depends on its initial state. However, when stochastic noise is present, the system will escape from any local minimum and will eventually end up in the absolute free energy minimum. This picture is consistent with selection of the unique equilibrium state at very late times as in \cite{pomeau,elder}.

The temporal evolution of many driven out of equilibrium systems cannot be described in terms of a potential as
\begin{eqnarray}\label{eq:dynamics}
\frac{\partial\phi}{\partial t}=-\frac{\delta{\cal F}(\phi)}{\delta\phi}+\zeta
\end{eqnarray}
where $\zeta({\bf r},t)$ is a Gaussian distributed uncorrelated random noise with $\langle\zeta({\bf r},t)\rangle=0$ and $\langle\zeta({\bf r},t)\zeta(0,0)\rangle=2\epsilon\delta({\bf r})\delta(t)$. A system whose evolution is described by Eq. (\ref{eq:dynamics}) will reach the stationary state corresponding to the absolute minimum of the potential ${\cal F}(\phi)$ as $t\rightarrow\infty$, as in the Swift-Hohenberg equation \cite{swifthohenberg}. The stochastic noise $\zeta({\bf r},t)$ in Eq. (\ref{eq:dynamics}) is essential for the selection of a unique stationary state provided that the noise strength $\epsilon$ is not too large \cite{elder}. If we interpret the potential ${\cal F}(\phi)$ as the free energy, then $\epsilon\propto T$, the temperature of the system. As expected, when $\epsilon>\epsilon_{c}$, or $T>T_{c}$, no unique state is selected and the system is disordered \cite{elder}. Also, selection of a unique equilibrium state occurs only in the thermodynamic limit when the probability of the state is unity. In a finite system, it is well known that fluctuations or noise ensure that all possible states are visited with finite probability so there is no true selection. 

The role of noise is not well understood, and mostly ignored in non-equilibrium systems on the grounds that it is so small that its neglect is justified \cite{Cross&Hohenberg}. An important question that has been addressed both experimentally and theoretically, is the effect of noise on the stability of stationary states or patterns\cite{theoraticalwork1,theoraticalwork2,theoraticalwork3,theoraticalwork4, theoraticalwork5,Garciabook,Garcia1,Garcia2,Garcia3}. Pattern selection is a widely investigated phenomenon. Systems where this has been observed include  convective rolls in Rayleigh-B\'{e}nard instabilities \cite{RB1,RB2,RB3}, propagation of Taylor vortices in unstable Couette-Taylor flow \cite{Couette-Taylor}, electrohydrodynamic convection in nematic liquid crystals \cite{liquidcrystal1,liquidcrystal2}, directional solidification \cite{directsold1,directsold2,directsold3} and fingering instabilities in Hele-Shaw type experiments \cite{dai,tomokazu}. These studies show that fluctuations seem to play an important role and that they should not be ignored. In this paper we further examine this phenomenon in an attempt to obtain a better understanding of the effects of noise. 

\begin{figure}
\includegraphics[width=0.5\textwidth]{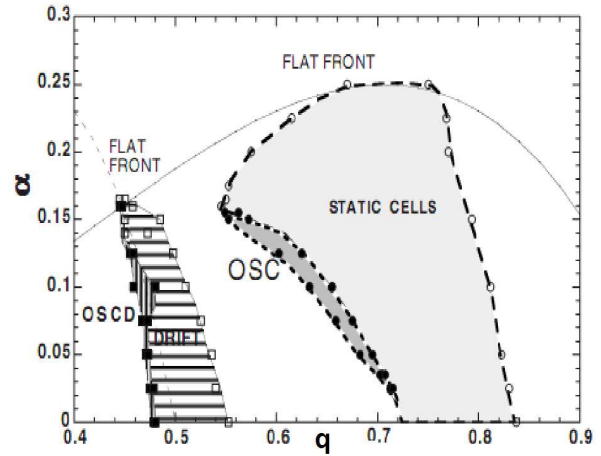}
\caption{Stability diagram of the primary and secondary states in the SKS equation. The domains of static cells, oscillating cells (OSC), drifting cells (DRIFT), and oscillating-drifting cells (OSCD), are bounded respectively by open circles, dark circles, open squares, and dark squares. In empty domains, a given state is unstable. The full line is the neutral curve of the mode of wave number q, and the dashed line for the mode 2q. The figure is taken from Ref. \cite{Brunet} (with permission).}
\label{figone}
\end{figure}

In dissipative systems, spatially periodic structures emerge as a result of a primary instability which develops when the control parameter exceeds its critical value \cite{eckhaus}. The periodic structures can themselves become unstable and develop secondary instabilities under certain perturbations of the primary pattern \cite{Cross&Hohenberg}. For example, the Eckhaus instability arises from the translational invariance of the periodic solution. As a result of this instability, the system undergoes a  change in the wavelength of the pattern. We consider the stabilized Kuramoto Sivashinsky (SKS) equation \cite{misbahvallance}
\begin{eqnarray}\label{eq1}
\partial_{t}u=-\alpha u-\mathbf{\nabla}^{2} u-\mathbf{\nabla}^{4} u + (\mathbf{\nabla} u)^{2},
\end{eqnarray}
where $u(x,t)$ is  a scalar function in one space dimension describing an interface profile. The control parameter, $\alpha>0$, is a stabilizing mechanism. The SKS equation is one of the simplest equations describing dissipative systems and is used to describe directional solidification \cite{misbahvallance} and  the Burton-Cabrera-Frank model of terrace growth \cite{bena}. Extensive studies \cite{misbahvallance, Brunet} reveal that this modest equation displays very rich dynamics and has a variety of secondary instabilities: (i) Eckhaus instability, (ii) period-halving of the cellular state, (iii) parity breaking, (iv) vacillating breathing and (v) oscillation with a spatial wavelength ``irrationally'' related to the basic one, as well as an abundance of tertiary instabilities. A linear stability analysis of Eq. (\ref{eq1}) about the trivial solution $u=0$ with $\delta u \sim e^{\lambda t+iqx}$ gives the dispersion relation $\lambda=-\alpha+q^2-q^4$ showing that a primary Turing bifurcation occurs at a critical value $\alpha_{c}=0.25$ and $q_{c}=\frac{1}{\sqrt{2}}$. For $\alpha\leq\alpha_{c}$, the trivial solution is unstable and modes in the band $1/2+\sqrt{1/4-\alpha}\geq q^{2}\geq 1/2-\sqrt{1/4-\alpha}$ grow exponentially. However, the nonlinearity in Eq. (\ref{eq1}) mixes these modes resulting in the emergence of stable periodic structures \cite{misbahvallance,eckhaus,Brunet}. The stability diagram of the SKS equation of Fig. (\ref{figone}) shows the neutral curve and the different regions of secondary bifurcations. 

In this work, we study the effect of noise on static cells in the stable region inside the Eckhaus stable band. The paper is organized as follows. In Section \ref{sec2}, we present the numerical simulations of the noisy SKS equation and the  results obtained. In Section \ref{sec3} we discuss the linear stability analysis of the deterministic noiseless SKS equation and the phase-diffusion coefficients of the steady states. We also present numerical computation of the phase-diffusion coefficients of the steady states of the deterministic SKS using  Auto \cite{sandstede,auto}, a software for continuation and bifurcations of a system of ordinary differential equations (ODE). We compare the results obtained from direct simulations of the noisy SKS equation with the computation of the phase-diffusion coefficient. Finally, in Section \ref{sec4} we draw some conclusions from our study.

\section{The Noisy Kuramoto-Sivashinsky Equation-Numerical Simulation}\label{sec2}

The Eckhaus stable band of stationary periodic states for the deterministic SKS equation is shown in Fig. (\ref{figone}) \cite{misbahvallance,eckhaus,Brunet}. All states inside this band are stable against small perturbations while those outside the band are unstable against long wavelength perturbations and decay to a state inside the band which depends on the exact form of the perturbation. Since any real system is subject to a variety of external and internal perturbations such as thermal fluctuations, vibrations from a slammed door, a heavy truck outside the laboratory, etc, it is of some interest to study the consequences of random noise on these stable stationary states. The simplest way to represent such effects is to add uncorrelated Gaussian distributed noise to the deterministic evolution equation. In this section, we investigate numerically the effects of additive Gaussian distributed white noise on the cellular stationary states of the SKS equation. The Langevin equation associated with Eq. (\ref{eq1}) is
\begin{eqnarray}\label{eq2}
\partial_{t}u=-\alpha u-\mathbf{\nabla}^{2} u-\mathbf{\nabla}^{4} u + (\mathbf{\nabla} u)^{2} +\zeta(x,t).
\end{eqnarray}
where $\zeta(x,t)$ is a Gaussian distributed noise with $\langle \zeta(x,t) \rangle =0$,  and  $\langle \zeta(x,t)\zeta(x',t')\rangle =2 \epsilon \delta (x-x') \delta (t-t')$ with $\langle\rangle$ denoting an average over the noise distribution. 

The noisy SKS equation of Eq. (\ref{eq2}) has been studied in the context of rough growth and morphological instabilities in various growth processes such as electrodeposition \cite{pastor,pastorrubio,buceta} and of the evolution of a surface undergoing ion sputtering \cite{chason,cuerno}. The focus of these earlier works is to understand the forced growth into a rough interface with emphasis on the interface width and its scaling properties. In this work, we focus the formation of a periodic stationary state and the the role of the stochastic noise in the selection of a unique stationary state rather than the scaling of the mesoscopic width of the interface. 

To perform numerical simulations, we discretize Eq. (\ref{eq2}) by defining $u_{i}^{j}=u(x_{i},t_{j})$ with $x_{i}=i\Delta x$, $t_{j}=j\Delta t$. The spatial grid of $N$ points is labelled by the subscript $1\leq i\leq N$ and the time grid of ${\cal N}$ points by the superscript $1\leq j\leq{\cal N}$. We impose periodic spatial boundary conditions $u^{j}_{i+N}=u^{j}_{i}$ and, since we have an initial value problem and a first order time derivative, specifying $u(x,t=0)$ completes the definition of the problem. The discretized Ito-Langevin equation corresponding to Eq. (\ref{eq2}) is \cite{Namiki} 
\begin{eqnarray}\label{eq3}
&&u_{i}^{j+1}=u_{i}^{j}+\Delta t F_{i}^{j}[u]+\sqrt{\frac{2\epsilon\Delta t}{\Delta x}}\eta_{i}^{j} \cr
&&F_{i}^{j}[u]=-\alpha u_{i}^{j}-\frac{1}{(\Delta x)^{2}}(u_{i+1}^{j}-2u_{i}^{j}+u_{i-1}^{j}) \cr
&&\quad -\frac{1}{(\Delta x)^{4}}(u_{i+2}^{j}-4u_{i+1}^{j}+6u_{i}^{j}-4u_{i-1}^{j}+u_{i-2}^{j}) \cr 
&& \quad +\frac{1}{4(\Delta x)^{2}}(u_{i+1}^{j}-u_{i-1}^{j})^{2}
\end{eqnarray}
where $\eta_{i}^{j}$ is a Gaussian white noise of unit variance with $\langle \eta_{i}^{j} \rangle =0$ and  $\langle \eta_{i}^{j}\eta_{k}^{l} \rangle =\delta_{ik}\delta_{jl}$. Because of the term in Eq. (\ref{eq3}), $(\nabla^{4}u)_{i}^{j}=(\Delta x)^{-4}(u_{i+2}^{j}-4u_{i+1}^{j}+6u_{i}^{j}-4u_{i-1}^{j}+u_{i-2}^{j})$ for $1\leq i\leq N$, it is convenient to define a grid of $N+4$ points labelled by $\alpha=-1,0,1\cdots N,N+1,N+2$ and $u_{\alpha+N}^{j}=u_{\alpha}^{j}$. 
 
We restrict our study to stationary periodic states by limiting the control parameter to $0.16\leq\alpha\leq 0.25$ \cite{misbahvallance,Brunet}. In the discrete system, the allowed wavenumbers are $q=2\pi N_{c}/L_{x}$ where $1\leq N_{c}\leq N$ is an integer corresponding to the number of cells in the pattern and $L_{x}=N\Delta x$ is the size of the system. Most of our simulations are done with $N=1024$ and grid spacing $\Delta x=0.50$ which is the largest value for which we could discern no difference in the pattern by changing $\Delta x\rightarrow\Delta x/2$. When we compare results for $\Delta x=1$ and $\Delta x=0.50$, the differences are visually very obvious but not for $\Delta x=0.50$ and $\Delta x=0.25$. The error due to the spatial discretization is ${\cal O}(\Delta x^{2})$. Because of the $\nabla^{4}u$ term in Eq. (\ref{eq2}), the Courant-Friedrichs-Lewy (CFL) condition determining the convergence of the discretized form in Eq. (\ref{eq3}) to the solution of the continuous PDE of Eq. (\ref{eq2}) is $\Delta t\leq C(\Delta x)^{4}$. Numerically, we find $C\approx 0.1331$ and we choose $\Delta t=0.006$ which satisfies the CFL condition when $\Delta x=0.50$. For $\Delta x=0.25$ the CFL condition requires a time step $\Delta t={\cal O}(10^{-5})$ which implies an order of magnitude greater computational time ${\cal N}\tau$ than for $\Delta x=0.50$ for the same real time $T={\cal N}\Delta t$. Here, $\tau\propto N$ is the time to perform a single update $t\rightarrow t+\Delta t$.

Direct simulations on systems of finite size, $L_{x}=N\Delta x$, show that noise destabilizes all steady states of the deterministic SKS equation of Eq. (\ref{eq1}), as expected. A truly stable stationary state can exist only in the thermodynamic limit $L_{x}\rightarrow\infty$ which implies that a finite size scaling analysis, similar to that used to study equilibrium phase transitions \cite{jylee}, might provide a more convincing case for state selection but this is beyond our ability. Thus, our conclusions from these simulations must rely on some operational definition of stability which we choose as the existence of the state in the presence of noise for at least some specified number of time steps. This measure allows us to compare the stability of different states. 

The Eckhaus band of spatially periodic stable states shrinks in the presence of noise. Even though we expect the stable band to shrink to a point, in the simulations we could only observe it to shrink to a narrow band in most cases. We were not able to further narrow down this region to a single state, because of limitations in both the computational time and allowed noise strength. In the absence of a clear definition of an appropriate noise strength, we limited ourselves to noise strengths for which we could still clearly observe a periodic pattern. The further from the boundaries of the Eckhaus band the fundamental wavenumber $q$ is chosen, the more stable against noise the periodic state becomes, so that a larger noise strength $\epsilon$, or, equivalently, a longer simulation time is required to destroy the state. As $\alpha$ is decreased, states inside the Eckhaus stable band become more stable.

\begin{figure*}
\centering
\includegraphics[width=1.0\textwidth]{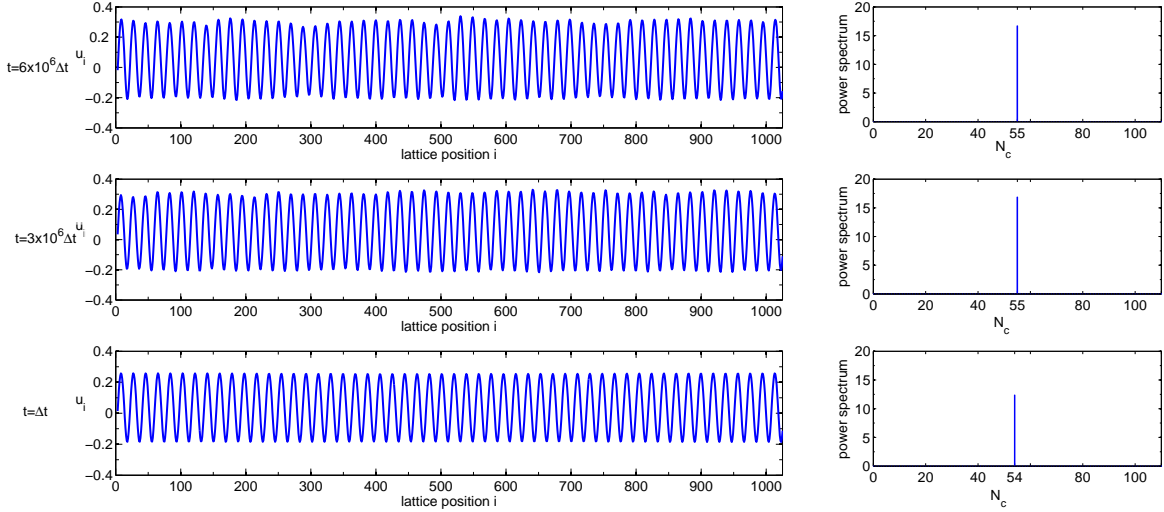}
\caption{\label{figtwo}Time evolution of an $N_c=54$ state at $\alpha=0.24$ in the presence of noise with $\epsilon=1.10^{-5}$.}
\end{figure*}

In the simulations, we take $\Delta x=0.50$, $\Delta t=0.006$ and $N=1024$ since these values allow for the largest real time $T={\cal N}\Delta t$ while satisfying the CFL condition. A few simulations with $N=2024$ were done with no change in the results. For $\alpha=0.24$, the band of stable wavenumbers is $53<N_{c}<61$ corresponding to $0.65< q_{s}< 0.74$. In the presence of noise with $\epsilon=1.10^{-5}$,  the initial  deterministic static steady state with $N_{c}=54$ evolves into the periodic state with $N_{c}=55$ shown in Fig. (\ref{figtwo}).
Similarly, the $N_{c}=60$ deterministic steady state transitions to the $N_{c}=59$ periodic state. How long a state of a particular periodicity survives before transitioning to another periodicity depends on the noise sequence generated. Also, the lower the noise strength $\epsilon$, the longer a particular periodic state survives before transitioning to a neighboring periodicity. For example, in a typical simulation with $\epsilon=2.10^{-4}$, the system transitioned from $N_{c}=59$ to $N_{c}=58$ after $7.10^6$ time steps. When $\epsilon=5.10^{-4}$, the same $N_{c}=59$ state evolved first into an $N_{c}=58$ state and then to a $N_{c}=57$ state in $3.10^6$  time steps. The $N_{c}=55$ state evolves into an $N_{c}=56$ state. The  $N_{c}=56$ and $N_{c}=58$ states in turn evolve into an $N_{c}=57$ or $q_{s}=0.6995$ state which appears to be the most stable state as it remains stable for $10^{8}$ time steps and noise strength up to $\epsilon=5.10^{-4}$.  
\begin{figure*}
\centering
\includegraphics[width=1.0\textwidth]{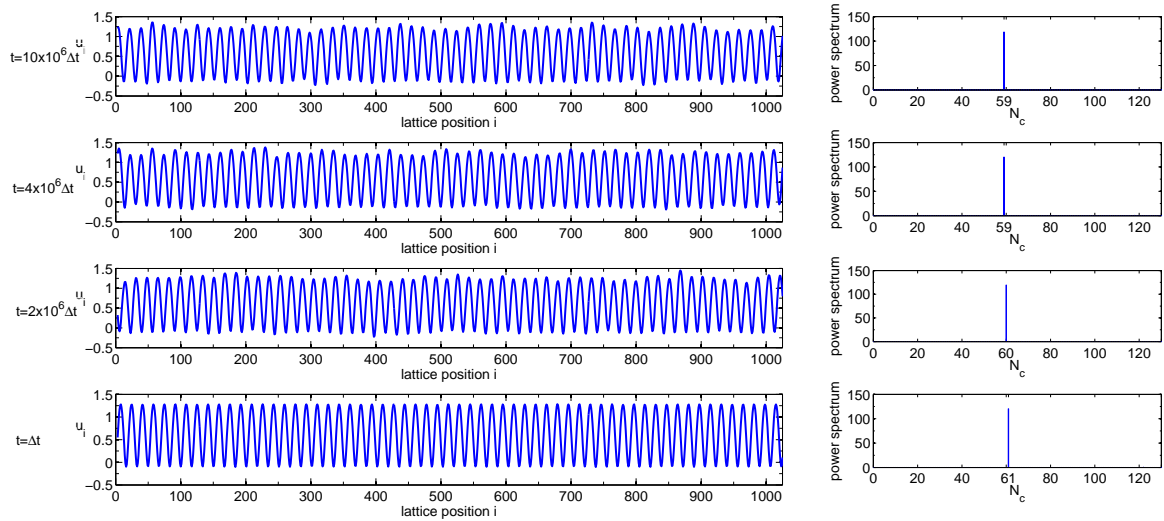}
\caption{\label{fig3}Time evolution of an $N_c=61$ state at $\alpha=0.2$ in the presence of noise with $\epsilon=5.10^{-4}$.}
\end{figure*}

As $\alpha$ is decreased, the states in the Eckhaus band become increasingly stable against noise and we could not shrink the Eckhaus band to a point but only to a narrower band. For $\alpha=0.2$, the Eckhaus band consists of  $0.58\leq q_{s}\leq 0.77$, corresponding to $48\leq N_{c}\leq 63$. In the presence of noise, states in the middle of the band survive. In Fig. (\ref{fig3}) is shown the evolution of an initial $N_{c}=61$ state to an $N_{c}=59$ state at $10^{7}$ time steps with noise strength $\epsilon=5.10^{-4}$. This state is not the most stable as six states with $53\leq N_{c}\leq 58$, corresponding to $0.65\leq q_{s}\leq 0.71$, remain stable for $1.10^{8}$ time steps against noise of strength $\epsilon\leq 2.10^{-3}$. 
\begin{figure*}
\centering
\includegraphics[width=1.0\textwidth]{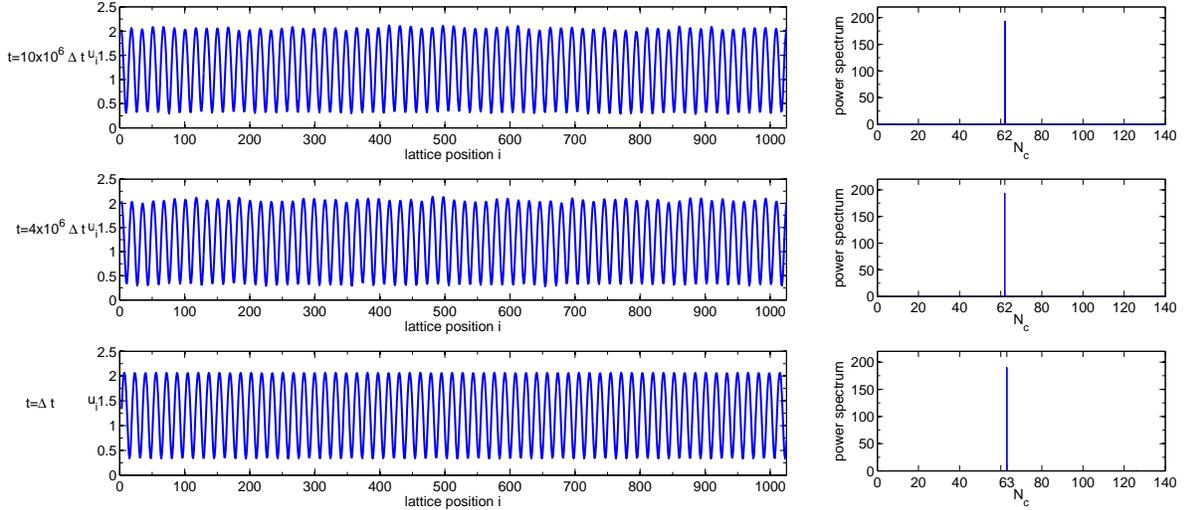}
\caption{\label{fig4}Time evolution of an $N_c=63$ state at $\alpha=0.17$ in the presence of noise with $\epsilon=1.10^{-4}$.}
\end{figure*}

\begin{figure*}
\centering
\includegraphics[width=1.0\textwidth]{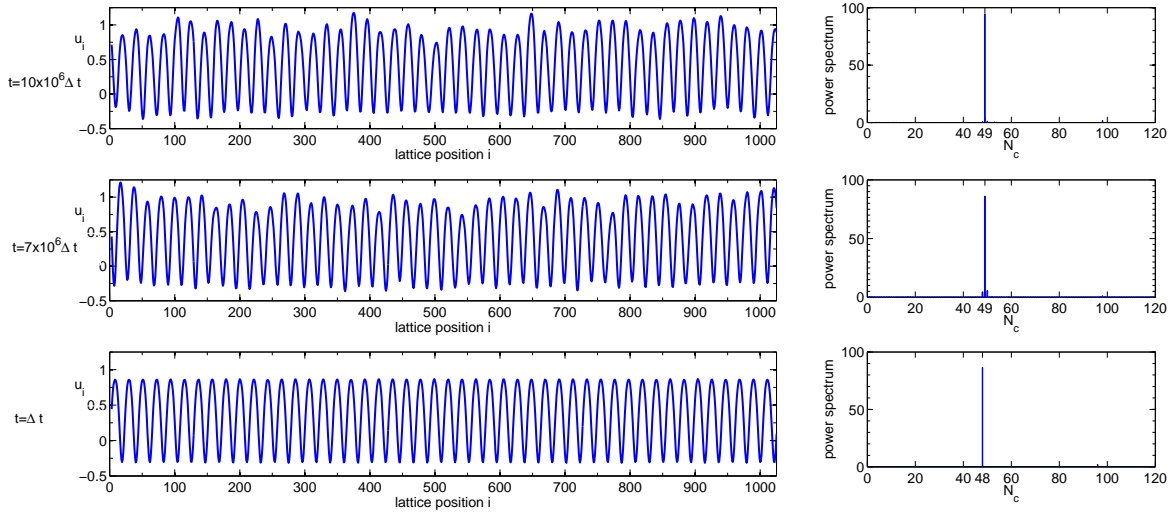}
\caption{\label{fig5}Time evolution of an $N_c=48$ state at $\alpha=0.17$ in the presence of noise with $\epsilon=5.10^{-4}$.}
\end{figure*}

For $\alpha=0.17$, Fig. (\ref{fig4}) shows the evolution of a periodic state with $N_{c}=63$ into a $N_{c}=62$ periodic state after $1.10^{7}$ time steps with noise strength $\epsilon=1.10^{-4}$. Fig. (\ref{fig5}) shows the evolution of a $N_{c}=48$ to a $N_{c}=49$ periodic state with a larger noise of strength $\epsilon=5.10^{-4}$. In this figure, small amplitudes of periodicities $N_{c}=48$ and $50$ with a main peak at $N_{c}=49$ can be seen in the snapshot of the pattern at $7.10^{6}$ time steps which have become much smaller by $10^{7}$ time steps when this simulation was terminated. In Fig. (\ref{fig5}), there is also a small second harmonic peak at $2N_{c}$ in the initial deterministic steady state with $N_{c}=48$ and in the $N_{c}=49$ state after $10^{7}$ time steps. The Eckhaus band at $\alpha=0.17$ contains periodicities $45\leq N_{c}\leq 65$ or $0.55\leq q_{s}\leq 0.80$ collapses into a narrower band of periodicities $53\leq N_{c}\leq 57$ or $0.65\leq q_{s}\leq 0.70$ in the presence of noise. All states in this narrow band remain stable against noise of strength $\epsilon=2.10^{-3}$ for $1.10^{8}$ time steps. For $\alpha=0.17$, states near the center of the Eckhaus band are very stable and destroying them by direct simulation of Eq. (\ref{eq2}) with noise strengths for which a periodic state clearly exists can take a {\it very}
long computational time. 
          
 \section{Phase Diffusion Coefficient}\label{sec3}

\subsection{Stability Analysis }
To  understand the destabilization of the static steady states in the presence of noise, we linearize the profile $u(x,t)$ about a periodic stationary state of period $L$, $u^{*}(x)=u^{*}(x+L)$
\begin{eqnarray}\label{eq5}
u(x,t)&=&u^{*}(x)+v(x,t) \cr
0&=&-(\alpha+\partial^{2}_{x}+\partial^{4}_{x})u^{*}+(\partial_{x}u^{*})^{2} \cr
v_t &=&[-\alpha-\partial_{x}^{2}-\partial_{x}^{4}+2 (\partial_{x} u^{*}) \partial_{x}]v\equiv{\cal L_{*}}v
\end{eqnarray}
where $v$ is an arbitrary function $v(x,t) \in \mathbb {C}^{n}$, and ${\cal L_{*}}$ is a linear operator which depends on the steady state solution $u^{*}(x)$. To check for the stability of the steady state, we compute the spectrum of ${\cal L_{*}}$, which reduces to solving the eigenvalue problem
\begin{eqnarray}\label{eq6}
{\cal L_{*}} v- \lambda v =0 
\end{eqnarray}
We define a Bloch wave operator
\begin{eqnarray}\label{eq7}
{\cal L_{\nu}}:= -\alpha - (\partial_{x}+\nu)^{2}-(\partial_{x}+ \nu)^{4}+2(\partial_{x} u^{*}) (\partial_{x} +\nu)
\end{eqnarray}
 where $\nu \in \mathbb{C}$. Eq. (\ref{eq6}) becomes 
\begin{eqnarray}\label{eq8}
[\alpha+(\partial_{x}+\nu)^{2}+(\partial_{x}+\nu)^{4}-2(\partial_{x}u^{*})(\partial_{x}+\nu)+\lambda]v=0
\end{eqnarray}
for some $v(x)=v(x+L)\in\mathbb{C}^{n}$ and some $\nu\in i[0,\frac{2\pi}{L})$. We focus our attention on the eigenvalue $\lambda_{o}$ defined such that $\Re{\lambda_{o}}\geq \Re{ \lambda_{j}}$ $ \forall j$, which determines the stability of the mode $u^{*}(x)$.

The Eckhaus instability is a long wavelength instability implying that $\nu=0$ is the important spatial eigenvalue \cite{Cross&Hohenberg}. Expanding $\lambda_{o}(\nu)$ about $\nu=0$, we obtain
\begin{eqnarray}\label{eq9}
\lambda_{o}(\nu)=\lambda_{o}(0)+ \nu \frac{d\lambda_{o}}{d\nu}|_{\nu=0}+\frac{\nu^{2}}{2}\frac{d^{2}\lambda_{o}}{d\nu^{2}}|_{\nu=0}+\cdots
\end{eqnarray}
We define the group velocity $c_{g}:=-\frac{d\lambda_{o}}{d\nu}|_{\nu=0}\in \mathbb{R}$  and the phase-diffusion coefficient $D:=\frac{1}{2} \frac{d^{2}\lambda_{o}}{d\nu^{2}}|_{\nu=0} \in \mathbb{R}$. For stable static cells, $\lambda_{o}(0)=0$ and $c_{g}=0$. Thus, $D(q)$ determines the stability of the periodic steady state with wavenumber $q$ \cite{misbahghazali,misbahpoliti}. 

\subsection{Computing the Phase Diffusion Coefficient}
\begin{figure}
\includegraphics[width=0.4\textwidth]{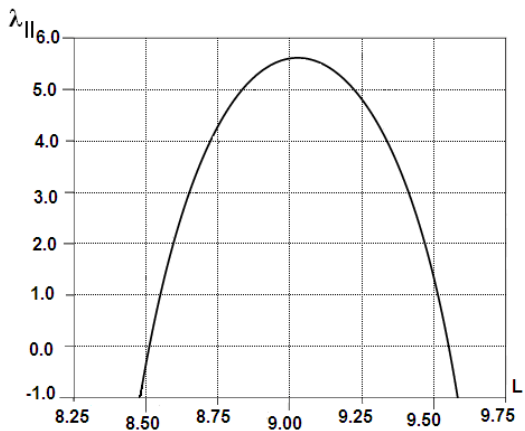}
\caption{\label{figthree} $\lambda_{||}=2D(q)$ as a function of the period $L=2\pi/q$ for $\alpha=0.24$. 
For $N=1024$ and $\Delta x=0.50$, $L=512/N_{c}$.} 
\end{figure}
\begin{figure}
\centering
\includegraphics[width=0.4\textwidth]{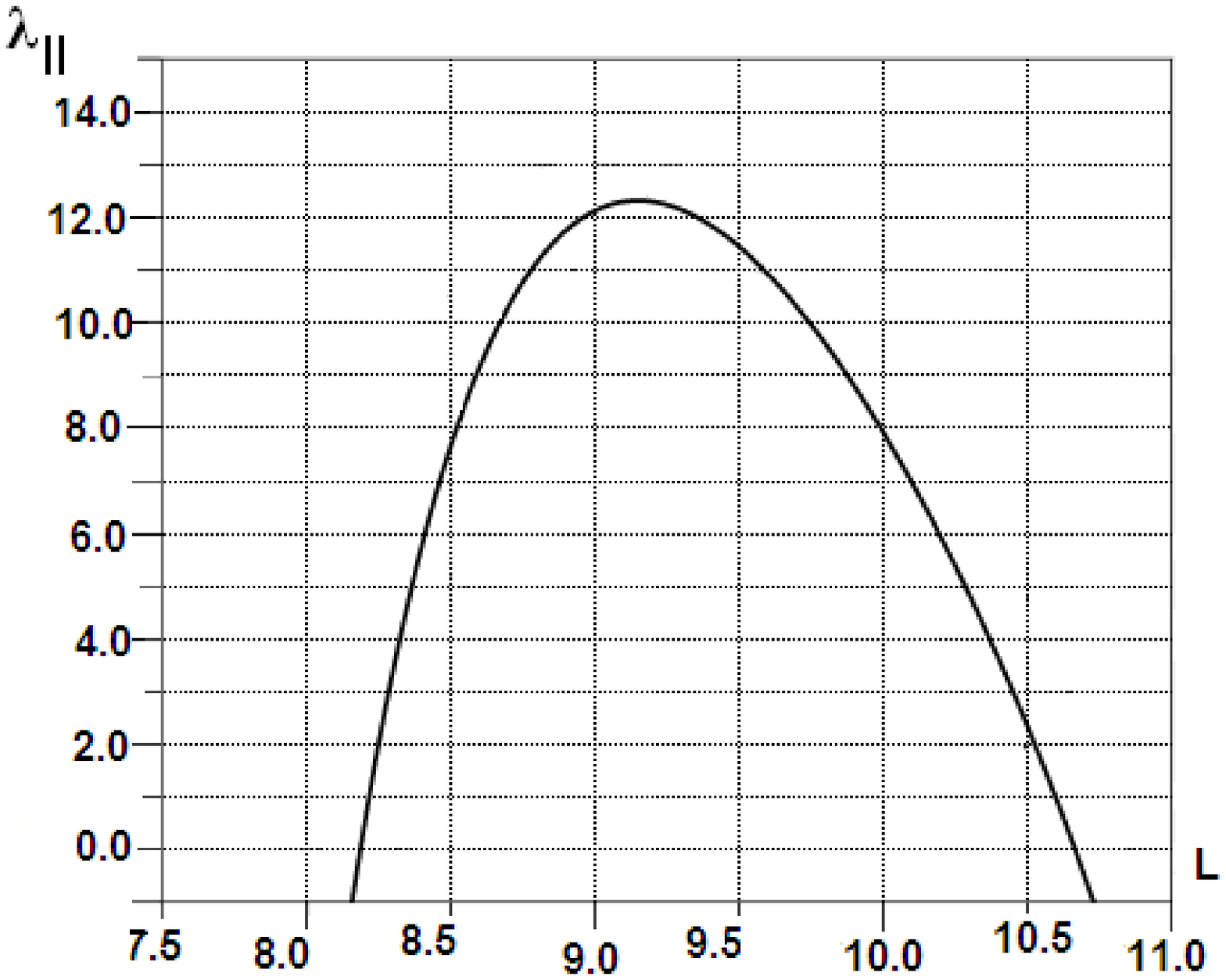}
\caption{\label{figfour} $\lambda_{||}=2D(q)$ as a function of the period L for $\alpha=0.2$. The wavenumber $q=2\pi/L$.}
\end{figure}
\begin{figure}
\includegraphics[width=0.4\textwidth]{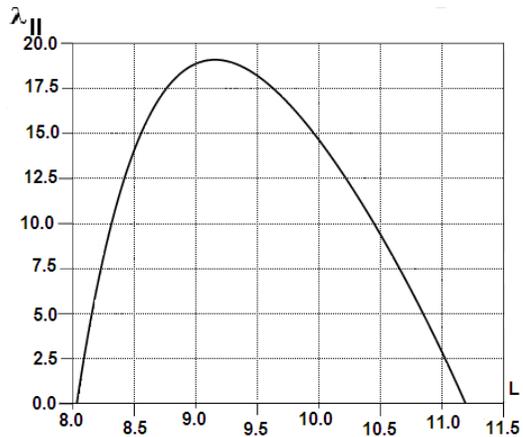}
\caption{\label{figfive} $\lambda_{||}=2D(q)$ as a function of the period L for $\alpha=0.17$. The wavenumber $q=2\pi/L$.}
\end{figure}

To find the eigenvalues as functions of $\nu$ and the diffusion coefficient $D(q)$, we use Auto \cite{sandstede,auto} which is a software for continuation and bifurcation of a system of autonomous ordinary differential equations (ODE). To use Auto, we rewrite the steady state equation associated with Eq. (\ref{eq1}) as a system of first order ODEs and, to get the correct number of parameters to solve the SKS equation, we add the term $c u_x$, where the parameter $c$ is the speed of the wave. This extra parameter is also used to check the results of the computations since it must always come out to be zero in our case. We write $U\equiv(u,u_{x},u_{xx},u_{xxx})$ which allows us to normalize the spatial period L to unity. We thus consider the boundary-value problem on the interval (0,1) \cite{sandstede}. As a first step, we must find the stationary solutions of the deterministic SKS equation within Auto as we use these solutions as input to the stability calculations. The stationary solutions $U$ obey the vector form of Eq. (\ref{eq5})
\begin{eqnarray}\label{eq10}
&&U_{x}=LF(U,c)U 
\end{eqnarray}
with the boundary and normalization conditions \cite{sandstede}
\begin{eqnarray}
&&U(1)=U(0)\nonumber \cr
&&\int_{0}^{1} \langle U'(x),U_{old}(x)-U(x) \rangle dx=0,
\end{eqnarray}
where $U_{old}(x)$ is the vector $U$ from the previous step in the iterative computational procedure, $U'=U_{x}$ and
\begin{eqnarray}
F(U,c)=\left(\begin{array}{cccc}
0&1&0&0\\
0&0&1&0\\
0&0&0&1\\
-\alpha&(u_{x}-c)&-1&0\end{array}\right).
\end{eqnarray}
Linearizing Eq. (\ref{eq10}) about a stationary solution $U^{*}$ by writing $U=U^{*}+V$ with $V=(v,v_{x},v_{xx},v_{xxx})$ gives the vector form of Eq. (\ref{eq8})
\begin{eqnarray}
\label{eq11}
&&V_{x}=L[A+\lambda B-\nu]V \cr 
&&A=\left(\begin{array}{cccc}
0&1&0&0\\
0&0&1&0\\
0&0&0&1\\
-\alpha&(2u^{*}_{x}-c)&-1&0\end{array}\right) \cr \cr
&&B=\left ( \begin{array}{cccc}
0&0&0&0\\
0&0&0&0\\
0&0&0&0\\
-1&0&0&0\end{array}\right)
 \end{eqnarray}
We supplement Eq. (\ref{eq11}) with the boundary and integral conditions
\begin{equation}                                                                                                      
\begin{array}{l}                                                                                                                                                                 
V(0)=V(1)\\                                                                                                                                                                      
\int_0^1 \langle V_{old}(x),V(x) \rangle dx=1 .                                                                                                                               
\end{array}
\end{equation}
Define $\lambda_{|}:=\frac{d\lambda_{o}}{d\nu}|_{\nu=0}$, and $\lambda_{||}:=\frac{d^{2}\lambda_{o}}{d\nu^{2}}|_{\nu=0}$ and denote $V_{x}$ by $V'$. Differentiating Eq. (\ref{eq11}) with respect to $\nu$ and evaluating the result for the most dangerous eigenvalues $\lambda=0=\nu$ gives \cite{sandstede}
\begin{eqnarray}
\label{eq12}
&&V'_|=L[AV_{|}+(\lambda_{|}B-1)V] \cr
&&V'_{||}=L[AV_{||}+2(\lambda_{|}B-1)V_|+\lambda_{||}BV]
\end{eqnarray}  
We impose boundary and integral conditions
\begin{eqnarray}\label{eq13}
&&\int_{0}^{1}\langle V(x),V_{|}(x)\rangle dx=0=\int_{0}^{1}\langle V(x),V_{||}(x)\rangle dx \cr
&&V_{|}(0)=V_{|}(1) \cr
&&V_{||}(0)=V_{||}(1)  
\end{eqnarray}
\begin{figure}
\includegraphics[width=0.5\textwidth]{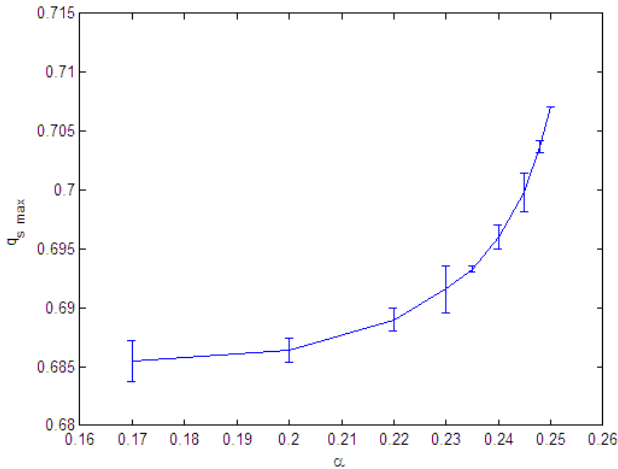}
\caption{\label{figsix} $q_{max}$, the wave number maximizing $D(q)$ as a function of $\alpha$. Error bars are largest possible errors.}
\end{figure}  
and obtain $\lambda_{|}$ and $\lambda_{||}$ by solving Eqs. (\ref{eq10}) to (\ref{eq13}) using Auto.
To guarantee convergence, we use a finite difference scheme to generate a steady state solution of the SKS equation, Eq. (\ref{eq10}), and use this as input for solving Eqs.  (\ref{eq11}-\ref{eq13}). The output of Auto consists of the eigenvalues $\lambda_{|}(L)$ and $\lambda_{||}(L)$ as functions of the period $L$ and the value of the wave speed $c$.  $\lambda_{|}$ and $c$ are always zero as expected for stationary periodic patterns. The non vanishing part of the output is $\lambda_{||}(L)$ as a function of the spatial period $L$ where $\lambda_{||}= 2D(q)$, the phase diffusion constant for fundamental wave number $q=2\pi/L$. $D(q) <0$ for periodic states outside the Eckhaus band indicating that these states are unstable against long wavelength perturbations. Following the standard interpretation, we identify the boundary of the Eckhaus band by the solution of $D(\alpha, q)=0$ and the stable band as the region $D(\alpha, q)>0$ \cite{Cross&Hohenberg,misbahvallance}. From Figs. (\ref{figthree}, \ref{figfour}, \ref{figfive}), we see that the maximum of $\lambda_{||}$ rises dramatically as $\alpha$ is decreased from $0.24$ to $0.17$. The wave number $q_{max}$  maximizing $\lambda_{||}(q)=2D(\alpha, q)$ is shown as a function of $\alpha$ in Fig. (\ref{figsix}). The error bars in Fig. (\ref{figsix}) are an overestimate as they represent the maximum possible uncertainty in an extrapolation from nearby points from Auto. We interpret these values of $q_{max}(\alpha)$ as the selected wave numbers as discussed in Section \ref{sec4}.

\section{Conclusions}\label{sec4}

The stability of the stationary periodic states of the deterministic SKS equation is studied (i) by simulation of the SKS equation with additive uncorrelated Gaussian distributed noise and (ii) by computing the decay rates of perturbations about these periodic states. Direct simulations of the noisy SKS equation show that the Eckhaus boundary separating unstable and stable spatially periodic states of the deterministic SKS equation collapses in the presence of additive Gaussian distributed stochastic noise and the Eckhaus stable band shrinks to a much narrower band. For $\alpha\alt\alpha_{c}=1/4$, the simulations show that the Eckhaus band shrinks to a point as one of the allowed values of $q$ is more stable than the others. However, as $\alpha_{c}- \alpha$ increases,  the periodic states in the Eckhaus stable band become more stable against noise and the simulations could verify only that the width of the stable band shrinks. These results agree with the picture obtained from the deterministic equation by computing the phase diffusion coefficient $D(q)$ of these spatially periodic states using Auto \cite{sandstede,auto}. 

We find that the value of the phase diffusion coefficient $D(q)$ is negative for periodic states outside the Eckhaus band, positive for those inside and zero for the states at the boundary.  This indicates that  states inside the Eckhaus band are stable against long wavelength perturbation. The phase diffusion coefficient seems to also determine the stability of the state against random fluctuations. The states with a higher value of the phase diffusion coefficient are more stable against noise. The maximum of the phase diffusion coefficient curve increases as $\alpha$ decreases. 

For $\alpha=0.24$, simulations of the noisy SKS equations show that the most stable state in the presence of noise is $N_{c}=57$ corresponding to $q=0.6995$ with parameters $N=1024$, $\Delta x=0.50$. This state  has the largest value of $D(q)$ of the states with $53\leq N_{c}\leq 61$ comprising the Eckhaus band for the $N=1024$, $\Delta x=0.50$, system used in our simulations. As $\alpha$ is decreased, the states inside the stable band become more stable against noise and, due to limitations in computational time and allowed noise strengths, we are able to narrow down this region to a single state for $\alpha=0.24$ only. However, the picture is consistent with that given by the phase diffusion coefficient despite the large $\Delta q=2\pi/(N\Delta x)=0.01227\cdots$ between allowed values of $q$ because of the periodic boundary conditions. The stable region in the presence of noise for each $\alpha$ corresponds to a region located about the maximum  of the phase diffusion coefficient curve for that particular value of $\alpha$. Also, from Figs. (\ref{figthree}, \ref{figfour}, \ref{figfive}), we see that the  maximum value of the phase diffusion coefficient increases from about $D\approx 2.75$ at $\alpha=0.24$ to $D\approx 9.5$ at $\alpha =0.17$, which is consistent with the observation from direct numerical simulations that the stability against noise of states inside the Eckhaus band increases as $\alpha$ decreases. 

We note from our results from the phase diffusion coefficient of Fig. (\ref{figsix}) that, as $\alpha$ ranges over the full range of stationary periodic states from $\alpha=0.25$ to $\alpha=0.17$, the periodicities vary over a very small range $0.707\geq q_{s}\geq 0.685$ corresponding to $58>N_{c}>56$ when $N=1024$ and $\Delta x=0.50$. To obtain a larger range of $N_{c}$, larger system sizes are necessary. Perhaps $N=2^{12}=4096$ is suitable but sizes like this will require extremely long runs and a large increase in the noise strength $\epsilon$. In turn, this will make identification of the periodicity of a state problematic. Taken together these extra difficulties seem to preclude numerical testing of our hypothesis with our presently available computational resources.  

\section*{Acknowledgments}
The authors thank G. Guralnik for access to his computational facilities and C. Pehlevan and C. Misbah for useful discussions.  D.O. thanks the School of Computational Sciences, KIAS, for support and hospitality when part of this work was done.

\end{document}